\documentclass[11pt]{article}

\usepackage[utf8]{inputenc}
\usepackage{graphicx}
\usepackage[english]{babel}
\usepackage{amsmath}
\usepackage{cite}
\usepackage{float}

\usepackage{authblk}
\usepackage[a4paper,no head,top=3cm,left=2.2cm,right=2.2cm,bottom=3cm]{geometry}
\begin{document}

\title{\bf{Electronic Instabilities of the AA-Honeycomb Bilayer}}
% \subtitle{subtitle}

%\maketitle

\author[1]{David S\'anchez de la Pe\~na \footnote {pena@physik.rwth-aachen.de}}

\author[2]{Michael M. Scherer}
\author[1]{Carsten Honerkamp}
\affil[1]{Institute for Theoretical Solid State Physics, RWTH Aachen University, D-52056, Germany,
and JARA Fundamentals of Future Information Technologies}
\affil[2]{Institute for Theoretical Physics, University of Heidelberg, D-69120, Germany}
%\shortauthors{David S\'anchez de la Pe\~na et al.}

%\twocolumn[
%  \begin{@twocolumnfalse}
    \maketitle
    \begin{abstract}
\normalsize
  We use a functional renormalization group approach to study the instabilities due to electron-electron interactions in a bilayer honeycomb
  lattice model with AA stacking, as it might be relevant for layered graphene with this structure. Starting with a tight-binding description for the four $\pi$-bands, 
  we integrate out the modes of the dispersion by successively lowering an infrared cutoff and determine the leading tendencies in the effective interactions. 
  The antiferromagnetic spin-density wave is an expected instability for dominant local repulsion among the electrons, but for nonlocal interaction terms also 
  other instabilities occur. We discuss the phase diagrams depending on the model parameters. We compare our results to single-layer graphene and the more common 
  AB-stacked bilayer, both qualitatively and quantitatively. \footnote {This is the pre-peer reviewed version of the following article: http://dx.doi.org/10.1002/andp.201400088}
   \end{abstract}

%  \end{@twocolumnfalse}]

%\shortabstract
%\begin{document}
%\maketitle

\section{INTRODUCTION}

Few-layer graphene systems are an object of current interest and debate (see e.g. \cite{Lau1,Lau2,Yacoby,Mayorov}). These systems can manifest different stacking orders, like simple hexagonal (AA), orthorhombic (AB) and rhombohedral (ABC) stackings. Regarding bilayer systems, the AB-stacked configuration has received the most attention so far, being the natural stacking in bulk graphite and energetically favoured over its AA counterpart. However, recent experiments show that AA stacking might be more common than previously thought \cite{Borysiuk,Lee}. The imaging of graphene bilayers using high resolution transmission electron microscopy of \cite{Liu} revealed that their samples are frequently AA-stacked. In practice, the obtainable bilayer samples are usually slightly twisted, exhibiting an alternating pattern of large regions with different stacking. The presence of AA-stacked regions has also been reported by scanning tunneling microscopy and photoemission measurements \cite{Lauffer,Kim}. After the experimental realizations, further theoretical studies have followed \cite{Brey,Nori,Nori2,Hsu,Xu,Chiu}. 

Of course, if a new physical system is available for further exploration of its properties and possible uses, it is mandatory to understand the electronic groundstate. In this work we study the possible groundstates of the AA-stacked honeycomb bilayer using a functional renormalization group (fRG) approach, which goes beyond the often-used random phase approximation (RPA), mean-field and 'g-ology' analyses. Starting from a given set of short-ranged interactions, the method allows for an unbiased investigation of the competing instabilities arising in the effective low-energy theory. The dominant tendencies indicate which correlations should prevail at zero temperature. Available ab-initio interaction parameters for graphene \cite{Wehling} are included among the different sets of interaction parameters we use, hoping that they bring a realistic picture for the groundstate of AA-bilayer graphene.

\section{MODEL}

Our starting point is a tight binding model for the AA-bilayer at half-filling, whose free Hamiltonian includes intra- and inter-layer hoppings to nearest neighbors (n.n.). The intra-layer or in-plane hopping term reads

\begin{equation}
H_{l}^{\parallel}=-t\sum_{<i,j>,s} (b^{\dagger}_{l,i,s}a_{l,j,s} + \text{h.c.})
\end{equation}

\noindent where $l=1,2$ is the layer index, $s=\uparrow,\downarrow$ is the spin, and $a^{\dagger}_{i}$,$a_{i}$ and $b^{\dagger}_{i}$,$b_{i}$ create and annihilate electrons at site $i$ on sublattice A and B respectively. The inter-layer or perpendicular hopping term is written as

\begin{equation}
H^{\perp}=-t_{\perp}\sum_{i,s}(a^{\dagger}_{1,i,s} a_{2,i,s} + b^{\dagger}_{1,i,s} b_{2,i,s} + \text{h.c.})
\end{equation}

\noindent In the following, the n.n. hopping $t$ sets the energy unit, and the spacing between n.n. lattice sites is taken as unity. Unless otherwise noted, the inter-layer hopping will be $t_{\perp}=0.1t$, as estimated in the literature \cite{McCann}. The diagonalization of $H_{\text{free}}=\sum_{l} H_{l}^{\parallel}+H^{\perp}$ yields 4 energy bands, which are two copies of the single-layer dispersion separated in energy by $2t_{\perp}$. For $\mu = 0$, the Dirac points at the $K,K'$ corners of the Brillouin zone have energies $\pm t_{\perp}$, and the band crossing between the two low-energy bands forms approximate circles at the Fermi level. This perfect nesting situation, together with the non vanishing density of states, makes the system particularly unstable once interactions are included. 

The interaction part of the Hamiltonian includes an on-site Coulomb repulsion $U$, intra-layer nearest neighbor and next to nearest neighbor repulsion terms $V_{1}$ and $V_{2}$, and an inter-layer repulsion between adjacent sites $V_{\text{il}}$  

\begin{equation}
H_{\text{int}}  =  U \sum_{i,l} n_{l,i,\uparrow}n_{l,i,\downarrow} + V_{1} \sum _{<i,j>,l,s,s'} n_{l,i,s} n_{l,j,s'} + V_{2} \sum _{<<i,j>>,l,s,s'} n_{l,i,s} n_{l,j,s'} + V_{\text{il}} \sum _{i,s,s'} n_{1,i,s}n_{2,i,s'}
\end{equation}

\noindent Since we will be working in the basis where $H_{\text{free}}$ is diagonal, one has to perform the same unitary transformation from orbitals to bands for $H_{\text{int}}$. The transformation has an angular dependence around the $K$,$K'$ points. This additional $k$-dependence of the interactions in band basis is often called \textit{orbital makeup}.

\begin{figure}
\includegraphics[width=0.8\columnwidth]{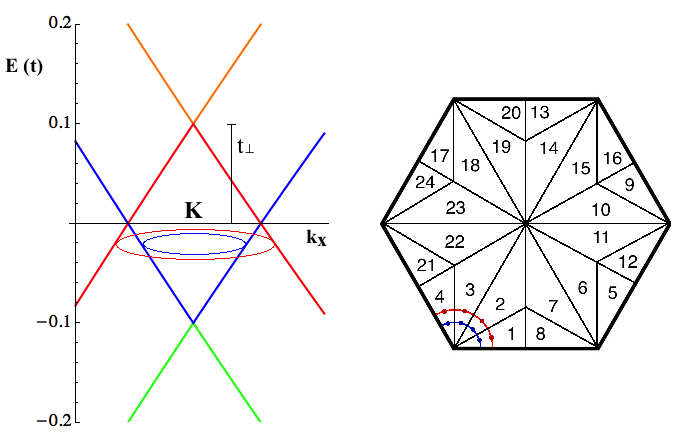}
\caption{\textit{Left panel:} Band structure of AA-bilayer graphene around a $K$ point with $k_y=0$. \textit{Right panel:} Patching of the Brillouin zone. Fermi surfaces are shown for small doping, with an exaggerated radius for illustration purposes. Dots denote the momenta representative for the patch's coupling value. At half-filling the two patchings lie on top of each other. The away bands are patched with small radii and are not shown here.}
\label{Fig1}
\end{figure}

\section{fRG METHOD}

The method we use is the one-loop, one-particle-irreducible (1PI) formalism of the fermionic functional renormalization group \cite{Metzner,Thomale}, in particular the scheme used in \cite{Trilayer}. 
The main procedure is to introduce a regulator depending on a scale parameter $\Lambda$ in the free part of the electron action, and to compute the variation respect to $\Lambda$, obtaining a functional flow equation for the 1PI vertices. 
Our regulator will be a momentum-shell cutoff at cutoff energy $\Lambda$. Setting the initial scale at the bandwidth ($\Lambda_0=W$) one has the bare action of the system as starting point for the RG flow, and as one integrates it down to $\Lambda \rightarrow 0$ the low-energy effective action is approached. The functional flow equation amounts to a coupled infinite hierarchy of vertex flow equations, thus to make calculations feasible the hierarchy is truncated after the four-point vertex $V_{\Lambda}$ (two-particle interaction), and the flow of the self-energy is neglected. 
This way we are left with one flow equation for the two-particle interaction only, whose solution amounts to an unbiased infinite-order summation of all possible combinations of one-loop particle-particle and particle-hole diagrams of second order in the interaction. The description of the four-point vertex is numerically implemented via a discretization of its quantum number dependences ($V_{\Lambda}=V_{\Lambda}(k_1,k_2;k_3,k_4)$ with $k_i$ including a Matsubara frequency $\omega_i$, a wavevector $\vec k_i$, a spin projection $s_i$ and either a band or orbital index $b_i$,$o_i$). Due to computing power limitations, frequency dependences are ignored, and being interested in groundstate properties the external ones are set to zero. The interaction preserves incoming spin projections, thus being spin independent. The remaining wavevector dependence is discretized in 24 patches as shown in figure \ref{Fig1}, with the coupling taking a constant value within each patch, and such value being evaluated at a representative point in the Fermi surface. 
Since the high-energy bands have no Fermi surface, they are patched with points at a smaller distance to the $K$,$K'$ points because that is where their extrema lie. Nevertheless, the precise choice does not alter the results. The discretization is done angularly around the $K$,$K'$ points since the orbital makeup varies only in the direction tangential to the Fermi surface. We also ran flows using a radially resolved scheme with 48 wavevector patches, where the radial split of each angular patch was done at a radius equidistant to the two Fermi surfaces. They yield the same results as the radially unresolved scheme. After discretization, for $N_b$ bands and $N$ patches one is left with a $N_b^4 \cdot N^3$ component coupling function $V_{\Lambda}$ (due to momentum conservation only three wavevectors are independent). Correspondingly, the vertex flow equation turns into a coupled set of $N_b^4 \cdot N^3$ differential equations.

\noindent This approximate scheme should be reliable up to intermediate interaction strengths. The approximations, however, do not allow the flow to be computed down to $\Lambda=0$, since some interaction components grow large in the process (called \textit{flow to strong coupling}). The flow must be stopped when the order of magnitude of the largest interaction component exceeds that of the bandwidth. This precise choice has no significant effect on the stopping scale, since the couplings diverge quickly as the instability is approached. The stopping scale provides an estimate for the critical scale $\Lambda_C$ at which a transition to a symmetry broken phase might take place. The interaction components growing large form sharp structures in wavevector space, signaling that the effective interaction becomes long-ranged. Moreover, the wavevector combinations at which such structures emerge allow for the identification of the corresponding instabilities. In this way, the fRG constitutes an unbiased tool for studying the interplay between different tendencies towards a symmetry broken state, instead of having to rely on some educated guess about the state that the system should adopt at low-energy (as is done in mean-field theory, for example).

\section{PHASE DIAGRAMS AND CRITICAL SCALES}

\begin{figure}[H]
\includegraphics[width=\columnwidth]{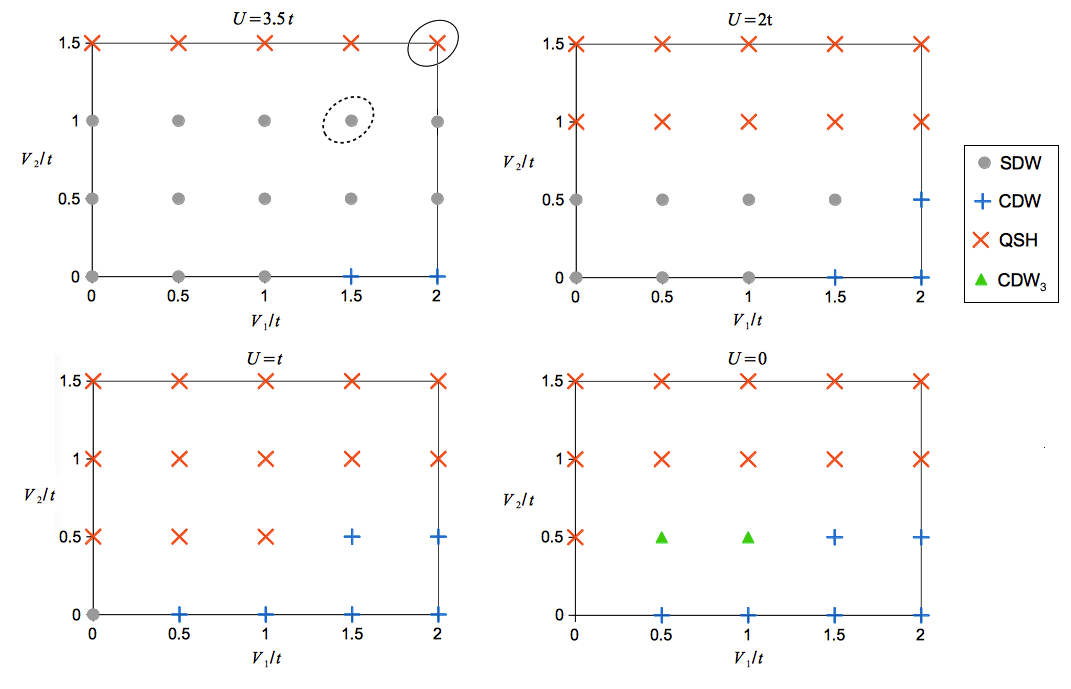}
\caption{Tentative phase diagrams for AA-bilayer graphene at half-filling, $T=0$ and $t_{\perp}=0.1t$. The cRPA parameters for graphene and graphite are encircled by solid and dashed lines respectively.}
\label{Fig2}
\end{figure}

In this work we present fRG results at $T=0$ for interaction parameters $U$,$V_1$ and $V_2$ up to the ab-initio values of \cite{Wehling}. As it turns out, $V_1$ and $V_{\text{il}}$ have the same effect on the instabilities of the system, thus the latter is omitted in the results shown. By identifying the leading instability for each parameter combination, a tentative phase diagram is obtained (Fig. \ref{Fig2}). The dominance among the coupling components changes gradually between different regimes, and so does the critical scale. Although having neglected self-energy effects a suppression of the scales due to finite quasiparticle lifetimes is not captured, the transition between these phases is expected to be of first order. We now proceed to describe the encountered instabilities.

\noindent \textit{Antiferromagnetic (AF) spin density wave (SDW) instability}. The tendency towards an AF-SDW state shows up in the fRG data as a divergence of interaction components with zero momentum transfer in the spin channel. The on-site interaction $U$ drives the flow towards this instability. The leading part of the effective interaction becomes 

\begin{equation}
H_{\text{AF}} = -\frac{1}{N} \sum_{o,o'}V_{o,o'}\epsilon_{o}\epsilon_{o'}\vec S^{o}\cdot \vec S^{o'}
\end{equation}

\noindent with $\vec S^{o}=\frac{1}{2} \sum_{\vec k,s,s'} \boldsymbol \sigma_{s,s'} c^{\dagger}_{\vec k,s,o} c_{\vec k,s',o}$. The orbital dependence is captured by the coefficients 
$V_{o,o'}>0$ and $\epsilon_{o} = +1$ for $o \in \{\text{A}_{1},\text{B}_{2}\}$, $\epsilon_{o} = -1$ for $o \in \{\text{A}_{2},\text{B}_{1}\}$. The interaction processes are attractive for intra-layer, intra-sublattice and inter-layer, inter-sublattice scatterings and repulsive for intra-layer, inter-sublattice and inter-layer, intra-sublattice scatterings. Since the effective interaction is infinitely long-ranged, a further mean-field treatment of $H_{\text{AF}}$ is justifiable. The mean-field decoupling reveals a layer antiferromagnet phase, where a net spin moment (e.g. $\uparrow$) is located on the $\text{A}_1$- and $\text{B}_2$-sublattices, and an opposite moment ($\downarrow$) on the $\text{B}_1$- and $\text{A}_2$-sublattices. In contrast with the AF-SDW state on the AB-bilayer studied in \cite{AB,Lang}, there is no difference in the sizes of the ordered spin moments between the sublattices, as all sites show the same connectivity. The spin quantization axis is not fixed. This phase opens a gap in the electronic spectrum.

\noindent \textit{Charge density wave (CDW) instability}. This instability takes place in the fRG flow as a divergence for interactions with zero momentum transfer in the density channel. This tendency is triggered by the intra- and inter-layer n.n. repulsion terms $V_1$ and $V_{\text{il}}$. The effective interaction reads 

\begin{equation}
H_{\text{CDW}} = -\frac{1}{N} \sum_{o,o'}V_{o,o'}\epsilon_{o}\epsilon_{o'} N^{o}N^{o'}
\end{equation}

\noindent with $V_{o,o'}>0$ and $N^{o}=\sum_{\vec k,s} c^{\dagger}_{\vec k,s,o} c_{\vec k,s,o}$. The orbital sign structure is the same as in the previous instability. This means there's an infinitely ranged attraction for sites on either same layer and sublattice or on a different layer and different sublattice, while there's repulsion for orbitals differing only in either layer or sublattice indices. After mean-field decoupling a gapped CDW phase arises, with a higher charge density in the $\text{A}_1$- and $\text{B}_2$-sublattices and a lower one in the $\text{A}_2$- and $\text{B}_1$-sublattices or vice versa.

\noindent \textit{Quantum Spin Hall (QSH) instability}. Another tendency can be observed in the spin channel for zero wavevector transfer, but with the distinctive feature of having an $f$-wave form factor. The $V_2$ interaction term is responsible for this instability. It corresponds to the effective Hamiltonian 

\begin{equation}
H_{\text{QSH}}=-\frac{1}{N} \sum_{o,o'}V_{o,o'}\epsilon_{o}\epsilon_{o'}\vec S^{o}_{f}\cdot \vec S^{o'}_{f}
\end{equation}

\noindent with $\vec S^{o}_{f}=\frac{1}{2} \sum_{\vec k,s,s'} f_{\vec k} \boldsymbol \sigma_{s,s'} c^{\dagger}_{\vec k,s,o} c_{\vec k,s',o}$ and $f_{\vec k}=\sin(k_x)-2\sin(\frac{k_x}{2})\cos(\frac{\sqrt{3}k_y}{2})$. The orbital sign structure is the same as before. However, interactions have now an additional sign structure that alternates between the $K$ and $K'$ points ($f$-wave modulation). In mean-field a pure imaginary Kane-Mele order parameter is induced, therefore the system enters a QSH phase. Having an even number of layers, this state supports an even number of helical edge modes and therefore they are not topologically protected. Note that the QSH state has also been looked for with more controlled numerical techniques for the single layer, albeit in finite-sized systems\cite{Daghofer}. No stable QSH  was found there, but at least enhanced correlations were detected. This raises the hope that the QSH state may actually be realizable in systems with higher density of states at the Fermi level. 

\noindent \textit{Three-sublattice CDW instability ($CDW_3$)}. A more exotic instability already found in previous fRG studies of few-layer graphene stacks\cite{AB,Trilayer} and in two other studies of the single layer\cite{Grushin,Daghofer} is present for the AA-bilayer as well, baptised as three-sublattice CDW due to the breaking of each sublattice intro three with different charge densities. It comes out as a divergence for interactions in the density channel with momentum transfer $\vec Q = \vec K - \vec K'$. The effective interaction in this case reads 

\begin{equation}
H_{\text{CDW}_3} = -\frac{1}{N} \sum_{o,o'}V_{o,o'}\epsilon_{o}\epsilon_{o'} (N^{o}_{\vec Q}N^{o'}_{-\vec Q}+N^{o}_{-\vec Q}N^{o'}_{\vec Q})
\end{equation}

\noindent with $N^{o}_{\vec Q}=\sum_{\vec k,s} c^{\dagger}_{\vec{k+Q},s,o} c_{\vec k,s,o}$, where $\vec k$ lies in the vicinity of the $K$,$K'$ points. We encounter the same orbital sign structure once again. The mean-field order parameter and electronic spectrum arising in this phase are described in \cite{AB,Trilayer}.

Having presented the phases emerging in this theoretical approach, we now comment on the possible implications for AA-bilayer graphene. The already mentioned ab-initio interaction parameters for graphene\cite{Wehling} were calculated through the constrained random phase approximation (cRPA). For these cRPA parameters the system is near the boundary between AF-SDW and QSH phases. The single-layer parameters lead to a QSH state, while the parameters for graphite place the system on the AF-SDW regime. Expecting the bilayer to take some intermediate values between these limiting cases, one is left at the region where the two instabilities compete. At this stage there cannot be a reliable prediction, since the precise outcome depends sensitively on the parameters. Interestingly, we found analogous instabilities as those already found in fRG analyses of the single-layer \cite{Singlelayer,Raghu}, AB-bilayer \cite{AB} and ABC-trilayer \cite{Trilayer}. The phase diagrams are remarkably similar too, therefore they are mainly determined by the intra-layer interaction physics plus the different densities of states of these systems. 
As in the previous fRG studies of few-layer graphene stacks \cite{Singlelayer,Raghu,AB,Trilayer}, no pairing instabilities dominate near half-filling and down to the scales considered, therefore no superconducting phase is observed. At a closer look, the main difference when comparing with the AB-bilayer and ABC-trilayer phase diagrams is a broader support for the QSH  in expense of the $\text{CDW}_3$ in the AA-bilayer.

\begin{figure}
\includegraphics[width=\columnwidth]{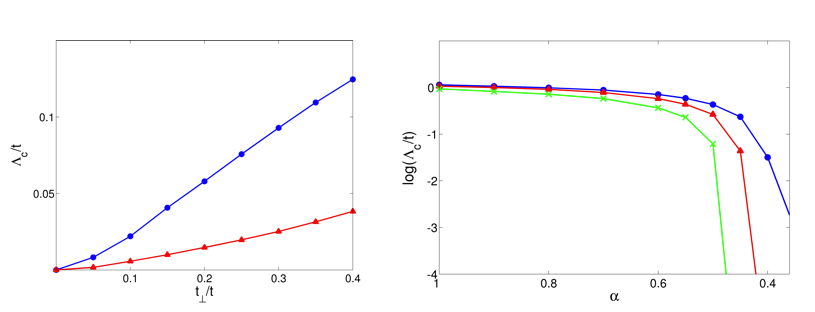}
\caption{\textit{Left panel:} fRG critical scale $\Lambda_c$ versus inter-layer hopping $t_{\perp}$ for the AB-bilayer (triangles) and AA-bilayer (circles), all in units of the intra-layer hopping $t$. \textit{Right panel:} Critical scale versus a rescaling of cRPA interaction parameters $\alpha$ for the single-layer (crosses), the AB-bilayer (triangles) and AA-bilayer (circles). }
\label{Fig3}
\end{figure}

The resulting critical scales are discussed next. Critical scales for the single-layer and AB-bilayer were also calculated for comparison. Figure \ref{Fig3} shows $\Lambda_C$ versus $t_{\perp}$ for both AA- and AB-bilayers with just a local on-site interaction, which was chosen to take the critical value for the single-layer within our scheme $U_C = 2.5t$. As expected, results for both stackings converge to the single-layer situation as $t_{\perp}\rightarrow 0$. The higher $\Lambda_C$ for the AA-bilayer is a consequence of the higher density of states at the Fermi level respect to the case with AB stacking. An analytic calculation using bands in a low-energy approximation reveals $\rho^{\text{AA}}_0=4\rho^{\text{AB}}_0$. For a more general choice of interaction parameters the scales can take values up to $\sim t$, a few orders of magnitude bigger than experimental results ($\sim 10^{-2}-10^{-3}t$ according to \cite{Lau1,Lau2}). This is particularly true for the cRPA parameters. It is nonetheless a rather ubiquitous discrepancy, as currently available many-body methods may greatly overestimate the critical scales when accounting for interaction effects in low-dimensional systems. As already hinted when presenting the band structure of this system, the perfect nesting between the two circular Fermi surfaces is the feature most responsible for such critical scales. In order to account for the approximations made in this fRG scheme, it can be argued that a reduction of the cRPA interaction parameters is necessary. Note that the graphene interaction parameters of the ab-initio method are well above the critical ones for the single-layer within our method. If we are to be consistent with the experimentally corroborated semimetallic nature of graphene, the interaction parameters have to be sub-critical for the single-layer. For that matter, we introduced a global rescaling parameter $\alpha$ for the cRPA interaction parameters ($U\rightarrow \alpha U$, $V_1 \rightarrow \alpha V_1$, $V_2 \rightarrow \alpha V_2$) and found that for values where the single-layer stays semimetallic, i.e. $\alpha \sim 0.5$, the bilayer critical scales drop down 2-3 orders of magnitude (Fig. \ref{Fig3}). This rescaling does not change the groundstate order, we observed the same instability for all $\alpha$. The critical scale of the AA-bilayer shows a slower decrease because the density of states stays constant in the interval $[-t_{\perp},t_{\perp}]$ around the Fermi level \cite{Tabert}, whereas the density of states in the other two configurations has a strong energy dependence.

In real systems, deviations from this simple model (e.g. impurities, lattice defects, etc...) probably play a role in keeping the critical scales low. Improvements in the theoretical analysis like including frequency dependence and self-energy effects should lower the critical scales too. A non-zero chemical potential breaks the Fermi surface nesting and hence suppresses the scales. We obtained results for small doping without qualitative changes but a reduction in the scales, albeit a mild one since the Fermi surfaces were still approximately nested. Further hopping terms do not have any major effect on the Fermi surface topology. An intra-layer next-to-nearest neighbor hopping $t'$ breaks the particle-hole symmetry, but leaves the band structure around the $K$ points intact. We performed calculations for $t'$ taking values up to $0.2t$ \cite{CastroNeto} and found again the same instabilities but smaller scales. The difference in this case is a consequence of the narrower set of bands starting to contribute later in the flow, as compared with the particle-hole symmetric case where they all contribute at the beginning of the flow already. Remote inter-layer hoppings produce an asymmetry between the two copies of the single-layer dispersion making up the AA-bilayer energy spectrum, while keeping the low-energy band structure unaffected. The band crossing is lifted slightly above the Fermi level, namely to energies $\sim 10^{-2}-10^{-3}t$ for hopping values in the literature \cite{Zhang,McCann}. As in the case of small doping, this should not alter any results qualitatively because divergences typically happen at larger scales, and the Fermi surfaces are very close to being perfectly nested. 

\section{DISCUSSION}

We presented here our extensive fRG calculations of the different groundstates arising in a honeycomb lattice model with AA stacking, as a model for AA-bilayer graphene. There is room for improvement of the method's exactness, but within its limitations, this approach surpasses conventional finite order perturbation theory and mean-field treatments. The robustness of the results should sustain confidence in the qualitative correctness of these theoretical outcomes. Unfortunately, uncertainties in the parameter values for the model and the approximations made in this scheme do not allow for a fully reliable quantitative description. Ab-initio values for the model parameters have been used, despite the necessity of a global rescaling of the interaction strengths. This was justified on the grounds of phenomenological consistency with the single-layer stability. The ab-initio interaction parameters place the system in a narrow competition between AF-SDW and QSH phases, where among other details the precise spatial profile of the interaction decides the winning tendency. Comparing the critical scales obtained, the AA-bilayer turns out to be more unstable than the AB-bilayer. Apart from the differences in the strengths and scales of possible instabilities, the nature of the leading correlations at low energy scales seems qualitatively the same for the single layer\cite{Singlelayer},  AA- and AB-bilayers \cite{AB}, and trilayers\cite{Trilayer} studied so far. Hence, the type of the leading correlation is determined by the in-plane physics, while the stacking dependent density of states decides if and at what energy scale the instability occurs. 

We thank Jie Yuan, Christoph Stampfer and Manuel Schmidt for discussions. We acknowledge support by DFG FOR 723, 912 and SPP 1459.

\bibliography{citations.bib}

\providecommand{\WileyBibTextsc}{}
\let\textsc\WileyBibTextsc
\providecommand{\othercit}{}
\providecommand{\jr}[1]{#1}
\providecommand{\etal}{~et~al.}


\begin{thebibliography}{[10]}

\bibitem{Lau1}% article
 \textsc{J.~{Velasco}},  \textsc{L.~{Jing}},  \textsc{W.~{Bao}},
  \textsc{Y.~{Lee}},  \textsc{P.~{Kratz}},  \textsc{V.~{Aji}},
  \textsc{M.~{Bockrath}},  \textsc{C.\,N. {Lau}},  \textsc{C.~{Varma}},
  \textsc{R.~{Stillwell}},  \textsc{D.~{Smirnov}},  \textsc{F.~{Zhang}},
  \textsc{J.~{Jung}},  and  \textsc{A.\,H. {MacDonald}} \jr{Nature
  Nanotechnology} \textbf{7}(March), 156--160 (2012).


\bibitem{Lau2}% article
 \textsc{W.~{Bao}},  \textsc{L.~{Jing}},  \textsc{J.~{Velasco}},
  \textsc{Y.~{Lee}},  \textsc{G.~{Liu}},  \textsc{D.~{Tran}},
  \textsc{B.~{Standley}},  \textsc{M.~{Aykol}},  \textsc{S.\,B. {Cronin}},
  \textsc{D.~{Smirnov}},  \textsc{M.~{Koshino}},  \textsc{E.~{McCann}},
  \textsc{M.~{Bockrath}},  and  \textsc{C.\,N. {Lau}} \jr{Nature Physics}
  \textbf{7}(December), 948--952 (2011).


\bibitem{Yacoby}% article
 \textsc{J.~Martin},  \textsc{B.\,E. Feldman},  \textsc{R.\,T. Weitz},
  \textsc{M.\,T. Allen},  and  \textsc{A.~Yacoby} \jr{Phys. Rev. Lett.}
  \textbf{105}(Dec), 256806 (2010).


\bibitem{Mayorov}% article
 \textsc{A.\,S. Mayorov},  \textsc{D.\,C. Elias},
  \textsc{M.~Mucha-Kruczynski},  \textsc{R.\,V. Gorbachev},
  \textsc{T.~Tudorovskiy},  \textsc{A.~Zhukov},  \textsc{S.\,V. Morozov},
  \textsc{M.\,I. Katsnelson},  \textsc{V.\,I. Falko},  \textsc{A.\,K. Geim},
  and  \textsc{K.\,S. Novoselov} \jr{Science} \textbf{333}(6044), 860--863
  (2011).


\bibitem{Borysiuk}% article
 \textsc{J.~{Borysiuk}},  \textsc{J.~{So{\l}tys}},  and  \textsc{J.~{Piechota}}
  \jr{Journal of Applied Physics} \textbf{109}(9), 093523 (2011).


\bibitem{Lee}% article
 \textsc{J.\,K. Lee},  \textsc{S.\,C. Lee},  \textsc{J.\,P. Ahn},
  \textsc{S.\,C. Kima},  \textsc{J.\,I. Wilson},  and  \textsc{P.~John} \jr{J.
  Chem. Phys.} \textbf{129}, 234709 (2008).


\bibitem{Liu}% article
 \textsc{Z.~Liu},  \textsc{K.~Suenaga},  \textsc{P.\,J.\,F. Harris},  and
  \textsc{S.~Iijima} \jr{Phys. Rev. Lett.} \textbf{102}(Jan), 015501 (2009).


\bibitem{Lauffer}% article
 \textsc{P.~Lauffer},  \textsc{K.\,V. Emtsev},  \textsc{R.~Graupner},
  \textsc{T.~Seyller},  \textsc{L.~Ley},  \textsc{S.\,A. Reshanov},  and
  \textsc{H.\,B. Weber} \jr{Phys. Rev. B} \textbf{77}(Apr), 155426 (2008).


\bibitem{Kim}% article
 \textsc{K.\,S. Kim},  \textsc{A.\,L. Walter},  \textsc{L.~Moreschini},
  \textsc{T.~Seyller},  \textsc{K.~Horn},  \textsc{E.~Rotenberg},  and
  \textsc{A.~Bostwick} \jr{Nat Mater} \textbf{12}(Oct), 887--892 (2013).


\bibitem{Brey}% article
 \textsc{L.~Brey} and  \textsc{H.\,A. Fertig} \jr{Phys. Rev. B}
  \textbf{87}(Mar), 115411 (2013).


\bibitem{Nori}% article
 \textsc{A.\,O. Sboychakov},  \textsc{A.\,V. Rozhkov},  \textsc{A.\,L.
  Rakhmanov},  and  \textsc{F.~Nori} \jr{Phys. Rev. B} \textbf{88}(Jul), 045409
  (2013).


\bibitem{Nori2}% article
 \textsc{A.\,O. Sboychakov},  \textsc{A.\,L. Rakhmanov},  \textsc{A.\,V.
  Rozhkov},  and  \textsc{F.~Nori} \jr{Phys. Rev. B} \textbf{87}(Mar), 121401
  (2013).


\bibitem{Hsu}% article
 \textsc{Y.\,F. {Hsu}} and  \textsc{G.\,Y. {Guo}} \jr{Phys. Rev. B}
  \textbf{82}(16), 165404 (2010).


\bibitem{Xu}% article
 \textsc{Y.~Xu},  \textsc{X.~Li},  and  \textsc{J.~Dong} \jr{Nanotechnology}
  \textbf{21}(6), 065711 (2010).


\bibitem{Chiu}% article
 \textsc{C.\,W. Chiu},  \textsc{S.\,H. Lee},  \textsc{S.\,C. Chen},
  \textsc{F.\,L. Shyu},  and  \textsc{M.\,F. Lin} \jr{New Journal of Physics}
  \textbf{12}(8), 083060 (2010).


\bibitem{Wehling}% article
 \textsc{T.\,O. Wehling},
  \textsc{E.~\ifmmode\,\mbox{\c{S}}\else\,\c{S}\fi{}a\ifmmode\,\mbox{\c{s}}\else\,\c{s}\fi{}\ifmmode\,\imath\,\else\,\i\,\fi{}o\ifmmode\,\breve{g}\else
  \u{g}\fi{}lu},  \textsc{C.~Friedrich},  \textsc{A.\,I. Lichtenstein},
  \textsc{M.\,I. Katsnelson},  and  \textsc{S.~Bl\"ugel} \jr{Phys. Rev. Lett.}
  \textbf{106}(Jun), 236805 (2011).


\bibitem{McCann}% article
 \textsc{E.~{McCann}} and  \textsc{M.~{Koshino}} \jr{Reports on Progress in
  Physics} \textbf{76}(5), 056503 (2013).


\bibitem{Metzner}% article
 \textsc{W.~Metzner},  \textsc{M.~Salmhofer},  \textsc{C.~Honerkamp},
  \textsc{V.~Meden},  and  \textsc{K.~Sch\"onhammer} \jr{Rev. Mod. Phys.}
  \textbf{84}(Mar), 299--352 (2012).


\bibitem{Thomale}% article
 \textsc{C.~{Platt}},  \textsc{W.~{Hanke}},  and  \textsc{R.~{Thomale}}
  \jr{Advances in Physics} \textbf{62}(November), 453--562 (2013).


\bibitem{Trilayer}% article
 \textsc{M.\,M. Scherer},  \textsc{S.~Uebelacker},  \textsc{D.\,D. Scherer},
  and  \textsc{C.~Honerkamp} \jr{Phys. Rev. B} \textbf{86}(Oct), 155415 (2012).


\bibitem{AB}% article
 \textsc{M.\,M. Scherer},  \textsc{S.~Uebelacker},  and  \textsc{C.~Honerkamp}
  \jr{Phys. Rev. B} \textbf{85}(Jun), 235408 (2012).


\bibitem{Lang}% article
 \textsc{T.\,C. {Lang}},  \textsc{Z.\,Y. {Meng}},  \textsc{M.\,M. {Scherer}},
  \textsc{S.~{Uebelacker}},  \textsc{F.\,F. {Assaad}},
  \textsc{A.~{Muramatsu}},  \textsc{C.~{Honerkamp}},  and  \textsc{S.~{Wessel}}
  \jr{Physical Review Letters} \textbf{109}(12), 126402 (2012).


\bibitem{Daghofer}% article
 \textsc{M.~Daghofer} and  \textsc{M.~Hohenadler} \jr{Phys. Rev. B}
  \textbf{89}(Jan), 035103 (2014).


\bibitem{Grushin}% article
 \textsc{A.\,G. Grushin},  \textsc{E.\,V. Castro},  \textsc{A.~Cortijo},
  \textsc{F.~de~Juan},  \textsc{M.\,A.\,H. Vozmediano},  and
  \textsc{B.~Valenzuela} \jr{Phys. Rev. B} \textbf{87}(Feb), 085136 (2013).


\bibitem{Singlelayer}% article
 \textsc{C.~Honerkamp} \jr{Phys. Rev. Lett.} \textbf{100}(Apr), 146404 (2008).


\bibitem{Raghu}% article
 \textsc{S.~Raghu},  \textsc{X.\,L. Qi},  \textsc{C.~Honerkamp},  and
  \textsc{S.\,C. Zhang} \jr{Phys. Rev. Lett.} \textbf{100}(Apr), 156401 (2008).


\bibitem{Tabert}% article
 \textsc{C.\,J. Tabert} and  \textsc{E.\,J. Nicol} \jr{Phys. Rev. B}
  \textbf{86}(Aug), 075439 (2012).


\bibitem{CastroNeto}% article
 \textsc{A.\,H. Castro~Neto},  \textsc{F.~Guinea},  \textsc{N.\,M.\,R. Peres},
  \textsc{K.\,S. Novoselov},  and  \textsc{A.\,K. Geim} \jr{Rev. Mod. Phys.}
  \textbf{81}(Jan), 109--162 (2009).


\bibitem{Zhang}% article
 \textsc{L.\,M. Zhang},  \textsc{Z.\,Q. Li},  \textsc{D.\,N. Basov},
  \textsc{M.\,M. Fogler},  \textsc{Z.~Hao},  and  \textsc{M.\,C. Martin}
  \jr{Phys. Rev. B} \textbf{78}(Dec), 235408 (2008).


\end{thebibliography}

% \begin{figure*}
% %   \twocolcaption
% %   \sidecaption
%   \includegraphics[width=\textwidth]{}%
%   \caption{\label{} 
%   }
% \end{figure*}
% 
% \begin{table}
%   \begin{andptabular}[<table width>]{<column declaration>}{<caption>}%
%     <table contents>\\
%   \end{andptabular}
% \end{table}
% \begin{table}
%   \begin{andptabbox}[<table width>]{<caption>}%
%     <contents>
%   \end{andptabbox}
% \end{table}
% 
% \begin{table*}
%   \begin{andptabular}[<table width>]{<column declaration>}{<caption>}%
%     <table contents>\\
%   \end{andptabular}
% \end{table*}

\end{document}